# Digital Twins for Industry 4.0 in the 6G Era

**Bin Han[1], Mohammad Asif Habibi[1], Bjoern Richerzhagen[2], Kim Schindhelm[2], Florian Zeiger[2], Fabrizio Lamberti[3], Filippo Gabriele Pratticò[3], Karthik Upadhya[4], Charalampos Korovesis[5], Ioannis-Prodromos Belikaidis[5], Panagiotis Demestichas[5], Siyu Yuan[1], and Hans D. Schotten[1,6]**

[1]Division of Wireless Communications and Radio Positioning (WICON), RPTU Kaiserslautern-Landau, 67663 Kaiserslautern, Germany
[2]Siemens Technology, 81739 München, Germany
[3]Department of Control and Computer Engineering, Politecnico di Torino, 10129 Turin, Italy
[4]Nokia Bell Labs, Espoo - 02610, Finland
[5]WINGS ICT Solutions SA, 17121 Athens, Greece
[6]Research Department of Intelligent Networks (IN), German Research Center of Artificial Intelligence (DFKI), 67663 Kaiserslautern, Germany

CORRESPONDING AUTHOR: Bin Han (e-mail: bin.han@rptu.de).

IEEE OJAP encourages responsible authorship practices and the provision of information about the specific contribution of each author.

This work has been funded by the European Commission through the H2020 project Hexa-X (GA no. 101015956).

**ABSTRACT** Having the Fifth Generation (5G) mobile communication system recently rolled out in many countries, the wireless community is now setting its eyes on the next era of Sixth Generation (6G). Inheriting from 5G its focus on industrial use cases, 6G is envisaged to become the infrastructural backbone of future intelligent industry. Especially, a combination of 6G and the emerging technologies of Digital Twins (DT) will give impetus to the next evolution of Industry 4.0 (I4.0) systems. This article provides a survey in the research area of 6G-empowered industrial DT system. With a novel vision of 6G industrial DT ecosystem, this survey discusses the ambitions and potential applications of industrial DT in the 6G era, identifying the emerging challenges as well as the key enabling technologies. The introduced ecosystem is supposed to bridge the gaps between humans, machines, and the data infrastructure, and therewith enable numerous novel application scenarios.

## I. INTRODUCTION

### A. A BRIEF HISTORY OF THE EVOLUTION OF MOBILE NETWORKS

Over the past fifty years, there have been significant advancements in communication and networking technologies, resulting in the birth of five generations of mobile networks. While these networks have shown significant improvements in peak data rates, the evolution of these networks has undergone a far more fundamental metamorphosis of paradigms. The Second Generation (2G) replaced the analog system with digital transmission technologies, while the Third Generation (3G) established a hierarchical cell structure, and the Fourth Generation (4G) became the first pure IP-based packet switching mobile network [1]. The latest Fifth Generation (5G) mobile network is taking a dramatic shift to network function virtualization (NFV), implementing most of its functionalities in a software-defined manner, paving the way for 5G to apply network slicing [2].

This shift towards network slicing enables heterogeneous services to coexist on top of a shared network infrastructure by means of logically independent network slices and supports specific use cases with extreme requirements. Additionally, 5G is targeting peak data rates of 20 Gbps, which is ten times faster than its predecessor 4G, and aims to provide ultra-low latency and massive connectivity for Internet of Things (IoT) applications [1]. Overall, while the mobile network evolution has resulted in higher peak data rates, it has also fundamentally transformed the paradigms of the network architecture, resulting in the development of 5G networks that are more flexible, efficient, and adaptable to support a diverse range of applications and services.

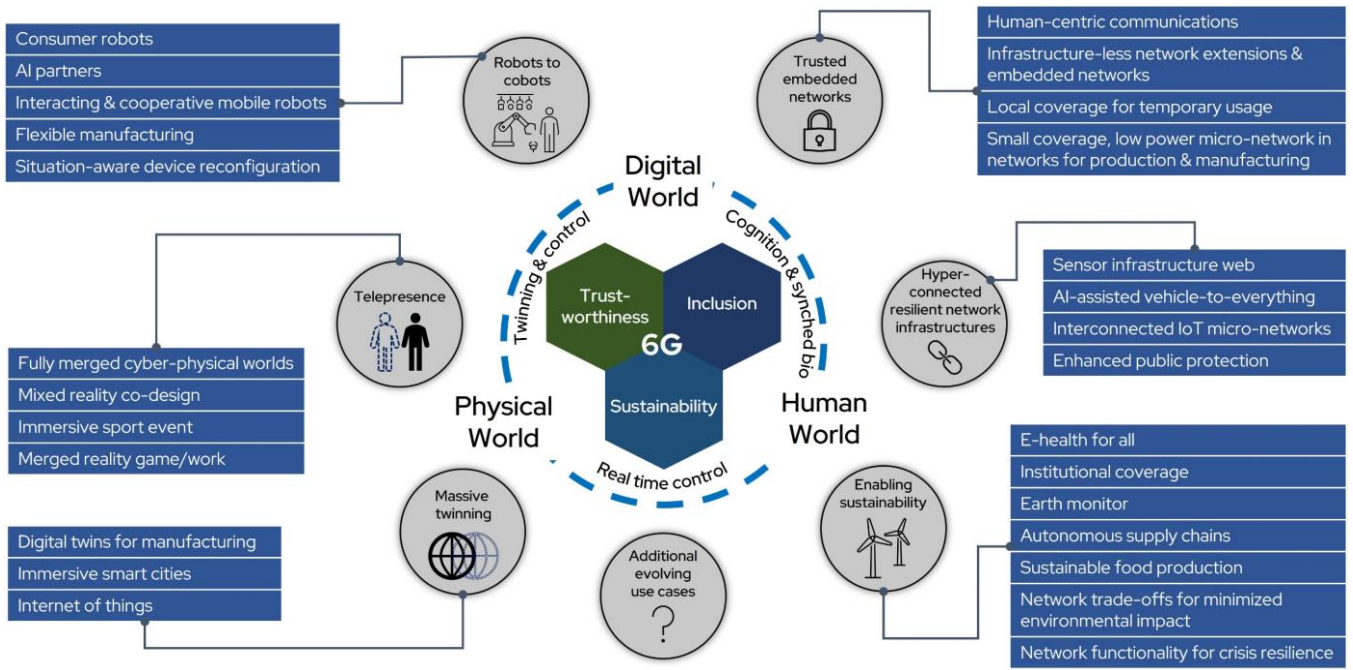

**FIGURE 1.** The Hexa-X vision of 6G use case families

### B. WHAT WILL 6G BE?
Beyond an incremental improvement of conventional key performance indicators (KPIs) such as throughput and latency, the next evolution of mobile networks is also expected to bring fundamental and revolutionary changes from the perspectives of technology principles, network architecture, and service paradigms, just as the previous evolutions did. Despite some hesitant voices [3], numerous visions for 6G have been proposed and are coming under the spotlight [4-9]. Generally, it is envisaged that 6G will be ubiquitously covering every corner of the planet [6, 8, 9], delivering not only data but also trustworthy intelligent services with extreme performance [8, 9]. Furthermore, growing concerns about the energy crisis and climate change are compelling people to prioritize sustainability as a key 6G value [10-12].

These opinions are fully reflected in the vision of the European Union's 6G flagship project, Hexa-X. As illustrated in Figure 1, Hexa-X envisages 6G to enable a plethora of emerging use cases that fall into six major families: sustainable development, massive twinning, telepresence, robots to cobots (collaborative robots), hyperconnected resilient network infrastructures, and trusted embedded networks. With these new use cases, 6G is supposed to connect the physical, digital, and human worlds with each other and drive the next evolution of cyber-physical systems. The convergence of the three worlds will not only further tighten the coupling between the physical and digital worlds, through ubiquitous connectivity and dependable control, but more importantly, will integrate humans seamlessly: humans and their relations shall be cognized and understood by the digital world, while also becoming capable of making real-time impact of the physical world.

Certainly, similar technical trends have been observed for more than a decade, especially in the consumer electronics industry, where personal context information about human users (e.g., health data or user preference profiles) is exploited to deliver personalized services [13]. However, restricted by the capabilities of telecommunication infrastructure, conventional context-aware mobile applications primarily focus on connecting the digital and human worlds, falling short of the requirements for ubiquitous connectivity and real-time tele-interaction. They are also frequently bonded to specific personal devices and consumer applications, limiting their flexibility and scalability in industrial environments. In the upcoming 6G era, contrarily, the inclusion of humans as the center of the system shall commonly take place in all kinds of applications and use scenarios, forming a global and immersive Internet-of-Everything-and-Everyone. This vision, once realized, will trigger a new revolution in the extant Industry 4.0 (I4.0) concept. Human beings shall integrate into the industrial system, not just as an exogenous subject controlling it, nor an object being served by it, but as an essential part of an indivisible, organic system.

In the light of this ambition, Hexa-X proposes 6G to be used for addressing six main critical challenges in the upcoming decade: 1) connecting diverse intelligent services across large-scale deployment; 2) establishing a single network of networks that aggregates multiple types of resources; 3) enhancing sustainability from environmental, economic, and social perspectives; 4) providing global service coverage; 5) enabling extreme experiences of services with ultra-high performance; and 6) ensuring a guaranteed trustworthiness in its services [14]. As an essential technical enabler of 6G,

Digital Twins (DT) will be deeply involved in the solution to each of these challenges [4,15].

I4.0 is commonly agreed to be characterized by the integration of cyber-physical systems, the Internet of Things (IoT), and cloud computing, among others, into manufacturing processes. DTs, which are virtual replicas of physical assets or systems that can be used to simulate, monitor, and optimize their behavior and performance, have shown a great potential in this scenario. Meanwhile, 6G wireless communication networks are expected to support a wide range of emerging applications and services, including those related to I4.0 and digital twins. The relationship between these three concepts is symbiotic, as I4.0 can benefit from the capabilities of DT technology and the 6G network, while DTs and 6G can leverage the data generated by I4.0 applications to improve their performance and effectiveness. We strongly believe that DTs, I4.0, and 6G are three interconnected concepts that play a crucial role in the ongoing digital transformation of various industries. In the remainder of this section, we will briefly review the concepts of I4.0 and DT, respectively, before diving into in-depth discussions on the envisioned impact of DT on I4.0 in the 6G era.

### C. INDUSTRY 4.0

The term I4.0 was coined to refer to a fourth industrial revolution as part of the "High-tech Strategy," a research agenda proposed by the German government [16]. Within this concept of a fourth industrial revolution, the impact of networking, DTs of machines and factories, distributed decision making, and (partially) autonomous operation – i.e., mostly originating in the world of Information Technology (IT) – on industry and the Operational Technology (OT) is summarized. This fourth industrial revolution is enabled by advances in communication technology, digitalization, and machine learning/artificial intelligence (among others), as discussed later in Section V.

As technologies and systems from both IT and OT need to work together to realize the vision of I4.0, standardization efforts affect both domains. The German industry founded the Standardization Council Industrie 4.0 (SCI 4.0) in 2016, originating from industry associations and platforms like Bitkom, DIN, DKE/VDE, VDMA and ZVEI. SCI 4.0 initiates and coordinates standards for digital production nationally and internationally, focusing on six thematic areas: (i) a Reference Architecture Model (RAMI 4.0), which is now published as IEC PAS 63088; (ii) an Asset Administration Shell (AAS); (iii) Artificial Intelligence; (iv) Security; (v) Sustainability and Digitalization; and (vi) Safety. A recent update on the state of standardization is given in [17], including references to relevant IEC and ISO standards. The link to communication networks as a building block of the I4.0 vision is reflected via activities in the IEEE for 802.11 and the 5G-ACIA for 5G and 6G systems. Appendix B in [17] contains a list of relevant national and international standardization bodies and initiatives.

### D. DIGITAL TWINS

The key idea of a DT is to create a digital (or virtual) replica of a physical system that is connected to it and is capable of making the associated information content accessible and actionable. While the concept has evolved over time, the fundamental idea of a DT can be traced back to an industry-targeted presentation at the University of Michigan about the creation of a center dedicated to Product Lifecycle Management (PLM) [18]. The term itself was not used (rather, the concept of "twinning" was referred to as "mirroring" operation), but all the elements of a DT, namely the data flow between the real world and its virtual counterpart (and vice versa), were already there, applied to the manufacturing domain. In [19], the conceptual model was applied in the astronautics and aerospace domains, and the result was finally referred to as a DT.

A DT can be used in many phases of a system's lifecycle. It can help in the prototyping phase by providing information necessary to describe, and then produce or develop, the physical representation of the system. More commonly, it acts as a virtual instance of the physical system, remaining connected to it throughout its lifetime. It can contain, among others, 3D models depicting the system's shape and structure, information about its components and their operation, as well as historical, current, and predicted data collected via embedded sensing mechanisms. DTs are typically used for monitoring, diagnostic, and prognostic purposes. They can also support the design, development, testing, and validation phases of any production and business process.

DTs are supported by a number of key enabling technologies and paradigms [20]. The first one is indeed represented by the IoT and communications in general. More specifically, great expectations are posed on 6G due to its forecasted ability to provide mobile connectivity to an enormous number of edge devices with high bandwidth and low latencies, which are critical for effective sensing and actuation in DTs [21]. Another prominent role will be played by artificial intelligence (AI), which will provide DTs with the level of intelligence required not just to process knowledge and let humans make data-driven decisions but also to make decisions autonomously [22]. Finally, extended reality (XR) will be critical for gaining insights into DTs by providing effective visualizations and supporting the implementation of telepresence-based solutions, allowing users to remotely operate on the DT [23].

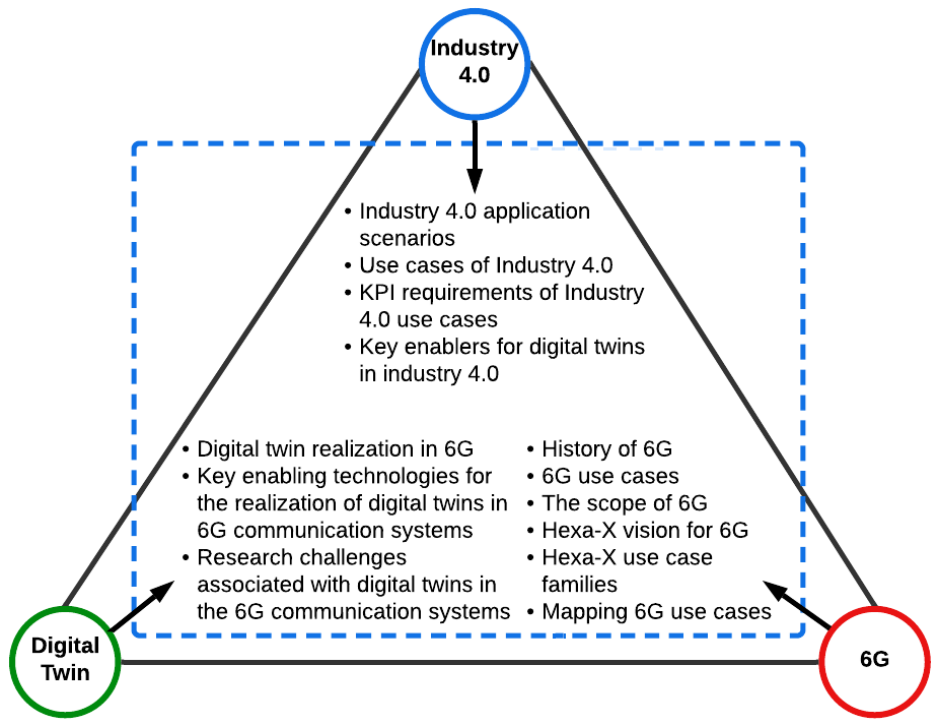

**FIGURE 2.** The relationship between the 6G, Industry 4.0, and Digital Twin. The scope of the paper is highlighted within the boundaries of the dashed blue box.

Although industry still represents the primary application domain for DTs [24], they are progressively being applied to a growing number of fields encompassing city management, construction, healthcare, and automotive, to name a few [20], with the paradigm extending beyond cyber-physical systems to create digital replicas of human beings, e.g., for wellbeing and coaching applications. Especially, the application of DT technologies can also enhance the performance and development of the 6G network itself by different means. More specifically, it enables efficient real-time monitoring and analysis of both the environments and the network status, assists dynamic network operation and management, and improves the training of AI models that can be used in 6G systems. Moreover, DT can be also used to ease and accelerate the assessment, development, and testing of technical enablers, novel services, and new deployment options of 6G [15].

### E. LITERATURE REVIEW AND THE SIGNIFICANCE OF THIS ARTICLE's CONTRIBUTIONS

At present, there are numerous survey papers available that provide an overview of DTs, I4.0, and their neighboring fields. To avoid duplicating the content of these survey and overview papers, we have provided a comprehensive list and comparative analysis of existing literature in our recently published survey paper [15]. Interested readers may refer to this paper and its references for more information on these topics.

However, in this paper, we aim to focus specifically on the digitization of the future I4.0 and the implications that DTs will have on various industries by using 6G communication networks and technologies. By doing so, we aim to address the research gaps in the literature and provide insights into the opportunities and challenges associated with the integration of DTs in the 6G era.

To enhance the paper's readability and ensure a coherent presentation of our findings, we have incorporated a comprehensive diagram outlining the overall framework in Figure 2. This diagram illustrates the interconnectedness of various sections, providing readers with an insightful visualization of the relationships between digital twins, Industry 4.0, and the 6G communication network. In addition, we have meticulously refined the section titles and introduced clear subheadings to align with our central argument. Furthermore, transition sentences have been introduced at the end of each section, facilitating a smooth flow of ideas, and enabling readers to grasp the logical progression of our research. Through these enhancements, we aim to present a unified narrative that showcases the relevance of digital twins in Industry 4.0 and their seamless integration with the state-of-the-art features of 6G technology. By offering improved readability and a more cohesive exploration of this emerging domain, we anticipate that our research will contribute significantly to the advancement of both digital twin applications and 6G communication networks.

### F. OVERVIEW AND STRUCTURE OF THE ARTICLE

The paper is structured into six sections, each focusing on a distinct aspect of the intersection of DTs and I4.0 in the 6G era. So far, Section I has introduced the concept of DT and its applications in Industry 4.0, followed by an overview of 6G communication networks and their potential to enable advanced DT applications. Section II presents a comprehensive background on DTs, including their ambitions, ecosystems, implications, and the perspectives of the various stakeholders and their roles in the development of DT technologies. Section III discusses the unique features of 6G communication networks that make them suitable for supporting a variety of DT applications in the forthcoming decade. In Section IV, we explore several emerging research challenges related to DTs in different industries that require sustained research efforts to fulfill I4.0 demands in the next decade. Section V delves into the key enabling technologies of DTs in the 6G era, while Section VI concludes the paper by summarizing the main contributions of the paper and discussing future research directions. For a better understanding and the convenience of the readers, we have summarized in Figure 3 the relevance of each of the three distinct technology domains (i.e., DT, I4.0, and 6G) to the various (sub)sections of the article. Compared to existing studies such like [25] and [26], the novelty of our work lies in its focus on the digitization of the future Industry 4.0 and the implications that DTs will have on various industries by utilizing 6G communication networks and technologies in the next decade.

| Section | 6G | I4.0 | DT |
|---|---|---|---|
| Section I: Introduction | | | |
| A | ● | | |
| B | ● | | |
| C | | ● | |
| D | | | ● |
| E | ● | ● | ● |
| F | | | |
| Section II: Ambitions | | | |
| | ● | ● | ● |
| Section III: Applications | | | |
| A | ● | ● | |
| B | | | ● |
| C | | ● | |
| D | | ● | |
| E | | ● | |
| F | | ● | |
| G | ● | | |
| H | ● | | |
| Section IV: Challenges | | | |
| A | | | ● |
| B | | | ● |
| C | ● | | |
| D | | | ● |
| E | ● | | |
| Section V: Enablers | | | |
| A | ● | | |
| B | | ● | ● |
| C | ● | | |
| D | | ● | |
| E | | | ● |
| F | ● | | |
| Section VI: Conclusion | | | |
| | ● | ● | ● |

**FIGURE 3. Overview of the article's structure regarding relevance to the three distinct technology domains**

## II. 6G INDUSTRIAL DTS: ECOSYSTEM AND AMBITIONS

In the envisaged *massive twinning* scenario, a DT can be created and maintained for everything (including both the assets and equipment of vertical industry and the network infrastructure) and everyone involved in the future 6G industrial scenario. The instantaneous observable states of the physical twins (PTs) shall be continuously monitored, and the measurements are uploaded to the computing nodes. Such computing nodes are usually located at the network edge, so that the latency can be minimized to keep the DTs as up-to-date as possible. Examples of such low-level state information are including but not limited to channel state information, computing resource availability, and user trajectory. By timely updating and archiving such data, industrial DTs are capable not only of monitoring the low-level status of massive machines and humans in real-time but also of estimating high-level context information that is typically unobservable, such as the physical or mental health condition of humans, the reliability of industrial equipment, or the possibility of malicious cyber-attacks. It will be possible to leverage this information in developing a 6G DT-based industrial ecosystem where the three worlds of data, machines, and humans converge into one. With ubiquitous, fast, and reliable connectivity for everything and everyone, the interconnection of physical entities and DTs will significantly expand the ways in which humans and things recognize and interact, leading to the following implications and potentials of DTs:

- **Supporting role of DTs in ubiquitous and collaborative telepresence:** To allow for virtual interaction between PTs over DTs, bidirectional data exchange between physical entities and their DTs must be executed in a timely and reliable manner. Despite early accomplishments to meet this requirement without 6G, a scalable and flexible deployment with numerous mobile devices involved is still waiting for 6G to support

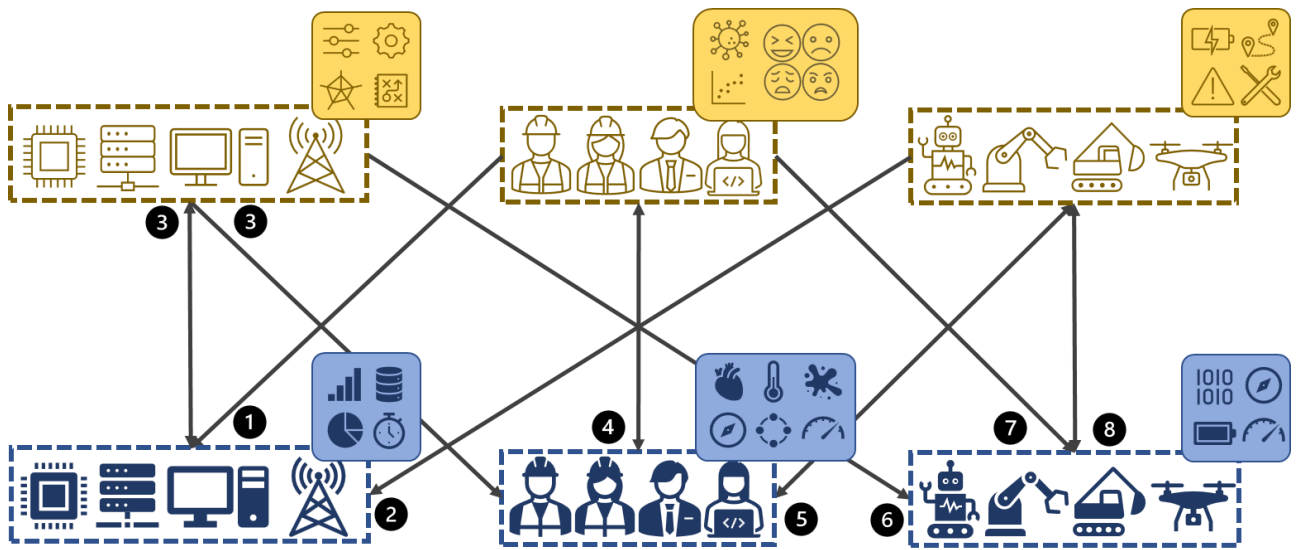

1: Human-aware networking
2: Ultra-flexible IoE networking
3: Network monitoring and analysis
4: Tele-collaboration
5: Remote monitoring and control
6: Network-aware smart factory
7: Cobots
8: Emergent intelligence

**FIGURE 4.** The industrial ecosystem of 6G DT, with arrows indicating the direction of information flow

simultaneous high throughput over massive data links [4]. Furthermore, with ubiquitous coverage and high accuracy in time synchronization expected from 6G, the same DT can simultaneously interact with multiple individuals in arbitrarily different locations without triggering conflicts or collisions between the operations. Thus, a remote collaboration that works as smoothly as on-site can be realized by means of telepresence, dramatically reducing the efforts and costs associated with spatial distance and access barriers.

- **Understanding the machines and environment:** By staying connected and synchronized with all elements of the physical world, a DT-enabled industrial system is capable of collecting vast amounts of data. The latest advances in machine learning (ML) and AI enable the extraction of reasoning models of different events hidden beneath the massive data. This will improve our understanding of industrial equipment and the physical environment, thereby enabling us to design, configure, operate, diagnose, and maintain them more effectively.

- **Potential of DTs in understanding the human participants:** While industrial equipment and the physical world mostly behave deterministically and can be precisely predicted using accurate models, human behavior is typically random and cannot be completely predicted. Therefore, human coexistence in industrial scenarios and human participation in industrial processes introduce significant uncertainty into the system, inevitably resulting in risks of service degradation and failure. Nevertheless, with massive data collected and analyzed, the DTs of human participants can create a statistical model of their behavior that can assist the industrial system in assessing and mitigating such risks.

Furthermore, using multi-dimensional status information obtained via advanced human-machine interfaces (HMIs), a human DT can recognize the statuses of its PT that are difficult or impossible to observe directly (e.g., emotion or fatigue), which can be exploited by machines to better comprehend humans in collaborative interactions.

- **Potential of DTs in sustainable industry:** On the one hand, with massive twinning, the status of all components and participants in industrial processes can be precisely monitored and jointly analyzed. This enables a thorough understanding of the unobservable patterns hidden beneath complex processes, such as carbon footprint and energy consumption along the process chain, that could potentially help in improving the energy efficiency and sustainability of the industry. On the other hand, novel solutions to green communication and energy-efficient computing are expected to ensure the sustainability of the industrial deployment of 6G massive twinning itself.

- **Benefiting more population:** DTs in I4.0 enable a broad range of remote interaction possibilities and human-in-the-loop use cases for remote maintenance and control or operations. Coupled with novel HMIs, Augmented Reality (AR), Virtual Reality (VR), and interaction with complex systems, this benefits remote workers and those with a handicap, increasing the potential reach of the respective jobs to a larger share of the population. With the capability to interact remotely with the DT, expert knowledge "as-a-service" is expected to become more accessible to smaller and medium-sized enterprises as well, creating potential for new players and smaller businesses to participate in the overall ecosystem.

Supported by these potentials, various novel application scenarios can be foreseen in such an ecosystem, where every physical entity is able to communicate and interact not only with its own DT, but also with the DTs of other entities. This concept is briefly exhibited in Figure 4, where the physical entities and their associated measurable data are illustrated in blue, while the DTs and their associated features are illustrated in red. The numbered arrows denote the information flow that support different novel application scenarios. For a better understanding of this paradigm shift, in the next section, we will selectively introduce several such industrial applications of 6G DT.

## III. APPLICATION SCENARIOS

In this section, we delve into various applications that are uniquely enabled by the seamless integration of the three key technology domains, i.e., 6G, DT, I4.0. Each application scenario discussed herein serves as a testament to the synergistic benefits that arise from this integration. For a comprehensive understanding of the requirements and challenges for these applications, we refer readers to our previous survey paper [15].

### A. HUMAN PRESENCE-AWARE URLLC

One industrial application scenario in which a human presence-aware approach becomes essential is providing ultra-reliable low-latency communication in a factory environment. The dominant type of traffic in an I4.0 factory environment is expected to be machine-to-machine traffic. To ensure low-latency communication across the entire factory floor, accurate modeling of rare events is required to avoid surprises. This is possible in future factories since they are semi-controlled environments with accurate position information available for fixed machinery as well as mobile user equipment (UEs) such as autonomous guided vehicles (AGVs). With accurate prior knowledge of the UEs' characteristics and capabilities, such as antenna beam patterns, number of antennas, etc., and location information and knowledge of the propagation environment, one can model the radio conditions between the UEs and the network using sophisticated methods such as ray tracing or full electromagnetic solvers.

However, when humans are present in the environment, their behavior becomes much more unpredictable and, unlike an AGV, whose movements are known or can be controlled, precise location information for a UE used by a human (e.g., a handheld device) is more difficult to obtain. In addition, the movement of such handheld UEs is inherently random and uncontrollable and complicates ensuring low-latency communication throughout the factory floor. Humans will also influence radio communication even if they are not active users, as the position of the limbs, clothing, and body type can all have an influence on the propagation environment.

Humans must therefore be modeled as objects in the DT, and their location and potential actions (movements, gestures, etc.) in a particular geographical region at a certain future time window must be considered when planning on how to use radio resources in that region over that time window. This modeling is critical for services having extremely high reliability targets. For instance, a simple hand wave can result in an outage for a service link that requires extremely low latency and has high reliability targets, and such movements have to be considered when planning the beams and frequency resources to be used. Similarly, enough redundancy and diversity can be prepared in advance if the likelihood of human blockage is high [27].

### B. MASSIVE TWINNING WITH HUMAN-IN-LOOP

As defined by Hexa-X, massive twinning is a use case family envisioned for the 6G era involving the massive use of DTs for the representation, interaction, and control of actions in the physical world. This use case family will be deployed in industrial environments, among others, enabling higher levels of production agility and more efficient interaction of production means to encompass to a greater extent the various processes and achieve the transfer of massive volumes of data, extreme performance, and reliability [28]. It relies on a fully synchronized and accurate digital representation of the physical and human worlds, with human expert presence being crucial for controlling, decision-making, corrective maintenance and operations adaptation.

Some example cases of massive twinning in manufacturing are the management of infrastructure resources with the use of various "what-if" scenarios through real time DTs or by handling a critical situation that may arise. Another application is the creation and testing of a new product in a virtual environment before physical manufacturing begins, as well as the detection and mitigation of anomalies in the real world through reconfigurations of the communication system and dynamic adaptations of the production process.

### C. FROM ROBOTS TO COBOTS

As anticipated, technology is changing the industrial domain, with new paradigms such as I4.0 progressively replacing traditional production processes. The role of robotics will indeed be key for the future of this new industrial revolution. In particular, research is focusing on three major directions [29]. One is represented by industrial robotics, intended as the use of robots that can serve a specific purpose and/or can be re-programmed to adapt to new production needs [30]. Another one is that of service robotics, which concerns the use of robots in diverse fields of human activity, including commercial and non-manufacturing-related tasks [31]. The third one is that of collaborative robotics, in which a shift is observed from robots to "cobots," i.e., machines designed to work in close cooperation with humans [32].

The latter direction is particularly attractive because it is expected to bring important advantages by combining the benefits deriving from the use of robots for automating repetitive jobs with the flexibility of manual operations performed by human operators [33].

Nevertheless, opportunities like that above are accompanied by a set of open challenges. The most critical ones are related to HMIs. As a matter of examples, well-established methods for robot programming are no more usable or not more effective with cobots and are being replaced by learning-by-demonstration methods based, e.g., on hand-guidance or other natural user interfaces [34]. Moreover, robots are no longer expected to operate in cells well-separated from humans' workspace; rather, robots and humans are expected to be co-located and to alternate in the execution of the tasks at hand. Hence, new ways of being aware of the collaborating peer's status and intentions are needed. This entails, among others, the use of sensors for collecting robots' and humans' poses; of AI-based algorithms for planning motion and preventing potential threats; as well as of solutions, e.g., based on AR and VR, to visualize relevant information from a distance (in virtual environments) or in the local user's field of view (overlapping to it) [35,36].

### D. EXTENDED REALITY (XR)

Artificial Intelligence and machine learning tools gathering information from robots and sensors can provide alerts, suggest changes or actions that the user can see and take control of whenever it is appropriate. In addition, important information on a specific part or tool should be presented in a simple and intuitive way to the user. For example, a user having AR, VR, or Mixed Reality (MR) enabled devices will be able to see information regarding the current input and output of an electric motor, its connections, and other characteristics without the need for a manual or measuring [37].

In I4.0, human machine interaction enhanced by XR will enable users to have more automated and supervisory roles, spending less time in planning, configuring, executing, and monitoring robotic tasks. New user roles will be introduced, from experts to non-expert users. Additionally, overloading users with unnecessary and difficult to understand data will be avoided, as well as recurring user's actions for confirmation for every task in robotic construction environment. Hence, the user's actions and decisions will be integrated when they bring remarkable results.

The human-machine interaction system should focus on three main interaction designs, supervision, notification, and control. In supervision, the user can monitor and supervise automated tasks. In notification, the system will notify the user of errors, exceptions, abnormal situations and any other valuable information and events. Finally, in control, the user will be able to take control of the system, reconfigure robots, networks and can even solve a problem with the use of teleoperation.

### E. NETWORK-AWARE DT FOR LOCAL INSIGHT

For I4.0 related use cases, as outlined in the previous section, a combined DT of the application and the utilized network, including its additional capabilities (e.g., localization, computation), is required to understand the implications of actions in the real world on the utilization of resources and performance of the network and vice versa. In particular, if DTs model mobile and moving assets, localization and sensing play a crucial role. Knowing the current location of mobile nodes and their surroundings enables the most efficient orchestration. Such a network-and capability-aware DT can then be used to locally optimize not only network operation but also application-specific aspects for a local (spatially confined) instantiation of the application in an I4.0 scenario [38]. One example is the planning of trajectories for AGVs in a flexible environment where humans are potentially crossing the respective pathways and the mobility of both, robots/AGVs and humans, impacts the received signal quality. If multiple options for communication exist (e.g., placement of computational capabilities on the end device or nearby edge servers, utilization of device-to-device (D2D) or device-to-infrastructure (D2I) communication, and potentially multiple access technologies), measuring and, consequently, modeling the individual options' impact on the performance perceived from the application perspective is required. In turn, this local insight can be utilized by the application to optimize AGV trajectories, taking network and application criteria into account.

### F. EMERGENT INTELLIGENCE

As an originally biological concept [39], emergent intelligence (EI) considers the intelligence of animals, including humans, as an emergent behavior, i.e., it spontaneously and emergently originates from a large number of interacting simple units. For example, human intelligence originates from the brain, which consists of massive simple neuron cells interconnected with each other. Another typical example is the collection intelligence of gregarious insects such as ants and bees, which have very simple behavioral patterns with every individual while exhibiting well-developed intelligence as colonies.

The phenomenon of EI has also inspired engineers to develop bionic intelligent approaches. It has been studied to a certain degree in the fields of complex systems and artificial intelligence. For instance, swarm algorithms have been developed over decades as a typical EI approach and an effective metaheuristic optimization tool. Some researchers also study genetic programming and genetic algorithms from the EI perspective. Further applications of EI have also been discussed in a wide range of areas, including logistics, resource allocation, job scheduling, and privacy, among others. Nevertheless, EI has generally not become a very significant research field so far.

According to [40], EI is distinguished from classical AI approaches mainly regarding the knowledge level and processing mechanism. It takes a decentralized and representation-specific processing approach, which relies on the interaction and local operations of independent agents to form an implicit global pattern at the macroscopic level, while none of the agents necessarily possesses knowledge about the system-level task. In contrast, AI integrates the explicit task knowledge into the problem solver, which operates globally with explicit knowledge of the target for the overall system.

**Table 1: KPI requirements for maintaining a DT in different use cases**

| | | Use-cases | | |
|---|---|---|---|---|
| | | Monitoring slowly varying parameters, e.g., temperature | Monitoring rapidly varying parameters, e.g., position for motion control, alarms, etc. | DT-related data transmission for offline processing |
| **KPIs** | Service availability [%] | >99.9 | ≤ 99.999999 | 99.9 |
| | Service reliability [%] | >99.9 | ≤ 99.999999 | 99.9 |
| | Latency [ms] | 100 | 0.1 | N/A |
| | Jitter [ms] | 10 | 0.01 | N/A |
| | Average Throughput [Gbps] | 0.01 | 10 | 10 |
| | Peak throughput [Gbps] | 0.01 | 100 | 100 |
| | Connection density [devices/sq. m] | 0.5 | 0.5 | 0.5 |
| | Safety | Critical | Critical | Low |
| | Integrity | Critical | Critical | High |
| | Maintainability | High | High | Low |

Gifted with these features, EI is showing competitiveness as a solution for spontaneous and decentralized intelligence, which is thirstily demanded by 6G to overcome the challenges of power efficiency, scalability, reliability, security, and privacy in delivering pervasive intelligence. On the other hand, EI generally requires a huge number of involved agents to achieve satisfactory performance, which implies a dense traffic load in wireless scenarios that is hard to handle. Fortunately, 6G is not only envisaged to provide ultra-dense access that globally interconnects everything, but also extensively deploy DTs, which can efficiently reduce the signaling overhead and wireless traffic load to enable EI. To put it briefly, 6G and EI match well together and can benefit from each other.

### G. DT-ASSISTED NETWORK SLICING ECOSYSTEM

Leveraging DT technology to emulate various network slice instances with different performance and configuration requirements can help a network operator lower total deployment cost, boost network slice planning and implementation, and enhance tenant experience [41]. To emulate the slicing ecosystem, we believe three DTs for each type of network slice instance are required: the underlying network infrastructure twin, the end-user's equipment twin, and the physical environment twin. The network infrastructure twin may emulate the underlying compute, storage, and networking resources of a mobile operator in order to provide both the operator and the slice owner with a visual representation of the required number of physical and virtual resources for a network slice instance. The end-user's equipment twin may be used to emulate the density of end-users, as well as their behavior and profile, served by a network slice instance. The physical environment twin may emulate spatial aspects (two-dimensional and/or three-dimensional representations) of the geographical area covered by the network slice to provide the slice owner and operator with a more complete understanding of the demographic data, space, and all physical entities surrounding the network sites. These three DTs of a network slice instance may interoperate or function independently. In either of these two cases, the network operator must make sure that the data from one twin does not get mixed up with the data from the other.

### H. SOFTWARE DEFINED RADIO; ENHANCED SPECTRAL AWARENESS; AND THE RADIO-AWARE DT

DTs can benefit overlay networks by providing the secondary network node with accurate information about the interference that it causes to the primary network. For instance, a DT of the primary network and the radio channel can aid in better utilization of the spectrum by a secondary network-node by utilizing the primary network's traffic pattern. If the secondary node determines from the DT that the interference from a secondary transmission to the primary network is small and does not violate the KPI targets for the primary network, the secondary network node can either act as a secondary transmitter or instruct a secondary UE to transmit on the resource that is in use by the primary network.

Similar benefits can also be seen when DTs are applied to facilitate new radio access technologies (RATs) deployed in a semi- or fully controlled IIoT environment. For example, a DT of the radio environment can assist with beam management for massive MIMO BSs at millimeter-wave (mmWave) and sub-THz wavelengths by reducing the number of beam candidates for a user node at a given location. The DT can also help reduce the overhead and latency associated with control signaling by reducing the need for measurements.

After discussing the application scenarios, now we are going to shift our focus to the challenges that arise when deploying DTs in 6G networks.

## IV. EMERGING CHALLENGES

This section highlights a number of critical research challenges that 6G communication systems must overcome to effectively deploy DTs. These challenges, which have been identified in [42], underscore the need for advancements in 6G technologies to meet the complex requirements of digital twinning.

### A. KEY PERFORMANCE INDICATORS

The KPIs for maintaining a DT depend on the use case in question. For instance, tracking rapidly changing parameters such as position and slowly/gradually varying parameters such as temperature will have fundamentally different requirements and KPIs. The relevant KPIs will also change based on how the communicated data is used inside the DT: for instance, communicating data for keeping a DT up to date will have different KPIs than communicating data that is used for offline processing, such as training a model. Therefore, important KPIs and target value ranges associated with them, based on [43], are described in Table 1 for different types of use cases. These KPIs underscore the technological demands that 6G must meet for effective real-time data synchronization and ultra-reliable communication.

### B. DEPENDABILITY AND SAFETY

The relation of dependability and DTs is two-folded. On the one hand, the information for DTs must be trustworthy so that deduced information can be considered as correct and used meaningfully in applications (dependability of the DTs). On the other hand, having a DT can support the requirement for dependability of systems (dependability supported by DT). Ensuring dependable DTs is a challenging task and encompasses many factors as, e.g., the availability, integrity, and reliability of data and systems must be ensured.

Data used to generate DTs must be trustworthy so that applications can generate meaningful outcomes: for example, when simulating certain problems/processes or when collecting near-real-time status quo information about factories to allow for predictive maintenance or quality of service optimization. Maintainability of DTs like modifications or repairs, even during runtime, must be considered as well. If DTs are used within safety applications to protect humans and ensure the absence of catastrophic consequences for the users and the environment, it becomes more challenging. Systems with safety requirements have to be assessed carefully, and the potential failure modes must be classified according to severity level. Standards and regulations exist depending on the field of application. The dependability and safety of DTs are directly linked to the capabilities of the 6G network, which must provide a reliable and secure communication infrastructure. If DTs play a role in such safety systems, the same rules and regulations are applicable.

If DTs of factories comprise the location of each machine, asset, and human within the factory, and if collisions between machines and humans are calculated via data gathered from 6G mobile networks, the position information, the calculation of collisions, and the automatic action instructions to avoid collisions must fulfill all safety requirements.

DTs can largely contribute to making dependable systems possible. For example, if predictions resulting from simulations based on DTs are always correct and the real-world scenarios behave just like the calculations based on the DTs, trust can be established in such systems and warnings can be considered reliable over time. The more physical world assets and processes can be modeled via DTs, the better the real-world processes and results can be calculated.

### C. SECURITY AND PRIVACY

As 6G will, with its ubiquitous coverage and massive twinning, connect everything and everyone while carrying massive amount of data that describes them comprehensively with more detail than ever before, it also raises concerns about security and privacy to an unprecedented level. The user data, which is consistently synchronized between the DT and its PT, can contain confidential or privacy-sensitive information. It must therefore be well protected, not only from unauthorized access and malicious operations by a non-trusted person or third party, but also from possible inappropriate exploitation by the trusted ones, such as the industrial verticals. For instance, the General Data Protection Regulation (GDPR) of the European Union prohibits a handful of kinds of data processing that can leak the user's identity or lead to discrimination, with only a few exceptions under very strict rules. How to exploit human-specific data in DT systems (especially those with DTs of humans) while staying aligned with such regulations must be taken into consideration when designing the HMIs. The heightened security and privacy concerns in a 6G-enabled DT ecosystem make it imperative for 6G technologies to offer advanced security features.

### D. SUSTAINABILITY

The envisaged evolution towards the future DT-driven industrial wireless networks and applications will lead to drastic changes in data traffic, information storage, and hardware infrastructure. Taking sustainability as one of the core values of 6G, measures must be taken in various aspects to tackle down the energy consumption issue in this process. For example, green solutions must be developed to collect data from numerous physical objects and maintain their DTs with high power efficiency. New network and data architectures must also be developed and standardized to enable the share of a DT by different stakeholders in different domains upon necessity, to mitigate unnecessary duplicates of the same object. Moreover, incremental solutions are called for to deploy DTs on the legacy infrastructure without a massive hardware replacement. These examples include the high computational power requirements, data storage needs, and communication demands. While these challenges can be significant, digital twins also have the potential to reduce energy consumption by enabling more efficient resource allocation and optimization. Overall, the examples discussed in this section demonstrate the need for careful consideration of energy consumption in the design and implementation of digital twin technology, which is highlighting the need for 6G to focus on energy-efficient solutions and resource optimization.

### E. INTERACTIONS BETWEEN DT NETWORK AND THE UNDERLYING INFRASTRUCTURE

While interacting with the underlying telecommunication infrastructure, the DT network performs the twining of an end-to-end 6G system into three layers: the data storage layer, the service mapping layer, and the management layer. These layers store various types of data, map the data to software analytics, and manage the twin throughout its lifetime, respectively. Nevertheless, such an ecosystem faces several challenges during this interaction.

The dataset(s) obtained by the storage layer may contain sensitive information about topologies, computation and communication resources, configuration parameters, and even the profile and behavior of customers. This may cause concerns for both the customer and the infrastructure provider. Hence, there is a need for a data collection mechanism that protects sensitive data while also creating an accurate twin of the underlying physical infrastructure.

The acquired data from the underlying infrastructure is in a variety of formats, features, and structures. Such raw data must be transformed into a unified format in order for the DT network to map the datasets onto software analytics tools and generate the twin of an infrastructure object. This data ingestion process delays the creation of the respective DTs. As a result, the DT network requires autonomous and AI-driven data ingestion mechanisms for unifying the input data.

On the management layer, the DT ecosystem interacts with a variety of physical objects manufactured by different vendors at various layers of the underlying infrastructure. Standard interfaces must be defined for a communication service provider to grant access to the underlying infrastructure via application programming interfaces (APIs) to a DT service provider. The intricate interactions between the DT network and the underlying 6G infrastructure highlight the need for 6G to provide advanced data collection and management mechanisms.

Having identified the challenges that shall be addressed by the future 6G technologies, in the next section we will discuss the technological advances that can serve as solutions.

## V. KEY ENABLING TECHNOLOGIES

To overcome the emerging challenges and open the door to new application scenarios for DT, 6G is supposed to introduce a variety of enabling technologies, which are briefly summarized in Table 2 and discussed in details below.

### A. RADIO ACCESS TECHNOLOGIES

In Section II, we have described some of the benefits that DTs offer for the deployment of new RATs. However, it is worth noting that this goes both ways where new RATs such as massive MIMO, cell-free MIMO, etc. and communication at mmWave and sub-THz provide a high-throughput, low-latency data pipe between the sensors and the DT that is essential for keeping the DT up to date.

From the RAT perspective, the exploitation of new spectrum, ranging from mmWave to sub-THz bands, will not only provide a huge bandwidth to support the deployment of massive twinning with sufficient data throughput and connection density, but also significantly shorten the minimal transmission time interval (TTI), which implies better time performance for more accurate synchronization among massive DTs. Furthermore, the short wavelengths in these new spectrum bands allow us to use compacter antennas and implement larger antenna arrays within a single device, making it realistic to achieve an extremely narrow beam width by deploying extra-large scaled massive MIMO (mMIMO). The spatial accuracy of beamforming is thereby increased, which not only enables physical-layer security for better protection of cyber-security and data privacy in DT applications but also effectively raises the spatial multiplexing gain and reduces cross-user interference, leading to a significant enhancement in energy efficiency. This gain can be even further amplified by introducing the so-called cell-free mMIMO, which also relies on perfect beamforming and allows multiple access points to jointly serve the same mobile device simultaneously. In addition, the recently emerging technologies of energy harvesting (EH) and simultaneous wireless information and power transfer (SWIPT) are expected to offer alternative options to replace batteries in 6G systems, for a more sustainable DT network.

### B. ARTIFICIAL INTELLIGENCE

Another technology that is expected to bring an essential added value to the DT paradigm, especially in the I4.0 perspective, is AI. A rich review of how AI is used in the field of industrial IoT context is provided in [44]. Through AI, the insights that are collected in the DT can be exploited by humans to make better decisions. AI can make DTs more intelligent, up to the point that they may be able to take decisions and perform actions on the physical entities they represent via so-called Intelligent Decision Support Systems (IDSS). For instance, by means of AI techniques, the DT of a plant can continuously monitor the status of machines and exploit gathered data to reconfigure in real time the underlying processes, e.g., to mitigate downtimes and bottlenecks [45]. AI capabilities and IDSS may also help in the logistics domain by allowing DTs to take data-driven decisions on planning and scheduling by considering, e.g., demand and distribution models. Actions like the above ones could support the implementation of optimization and control strategies targeted to improve efficiency and, ultimately, profitability [46]. Thanks also to forecasted advancements in mobile networks and edge computing capabilities, benefits of AI can be largely

VOLUME XX, XXXX 9

awaited also in the context of HMIs for DTs. A typical use case in this context is that of robotics where, e.g., computer vision is essential for the navigation of mobile robots [47] and the interaction with collaborative robots (or cobots). Another typical application of AI technique in this field is that of human-action recognition from images and data collected by other sensors (like depth cameras) to perform, e.g., trajectory forecasting and path planning for safety assurance in scenarios

**Table 2: Technical enablers for 6G-DT in I4.0 and their impacts**

| Technical enabler | | Addressed challenge(s) | Involved application scenario(s) |
|---|---|---|---|
| Category | Technologies | | |
| **RAT**[(1)] | mMIMO, extra-large mMIMO, cell-free mMIMO | Service availability & reliability, latency, jitter, throughput, connection density, sustainability, energy efficiency | Massive twinning with human-in-loop, Cobots, AR/VR |
| | New spectrum: mmWave and sub-THz | Service availability & reliability, latency, jitter, throughput, connection density, sustainability, security & privacy, energy efficiency | Massive twinning with human-in-loop, Cobots, AR/VR |
| | Energy harvesting and SWIPT | sustainability | Massive twinning, Cobots, AR/VR |
| **AI/ML** | IDSS for production management | Downtimes and bottlenecks | Monitor machines' status, reconfigure processes in real time |
| | IDSS for logistics | Resource underutilization, efficiency, profitability | Demand management and distribution model optimization |
| | Robotics, sensing | Mobile robot navigation, human-robot co-located collaboration, safety in production | Automated, human-aware manufacturing |
| **MEC** | Industrial campus network | Reliability / dependability, latency, security and privacy | All |
| **Sensing & Positioning** | Integrated Communication and Sensing, IRS | Efficient integration of communication, localization and sensing, Service availability & reliability, latency | Dependable DT |
| **HMI** | Sensing the users' status and the environment | Safety; benefiting more population | Human-presence-aware industry, cobots |
| | Multi-sensory feedback | benefiting more population | Cobots |
| **CoCoCoCo** | | Service availability and reliability, sustainability | Network aware DT for local insight |

involving the operation of co-located human and robotic agents [48, 49].

### C. MULTI-ACCESS EDGE COMPUTING
The deployment of modern AI algorithms generally requires rich computing power, and this is where Multi-Access Edge Computing (MEC) plays its key role in 6G industrial DT systems. Specifically, by bringing computing capabilities to the edge of network, the industrial edge networking, with its associated concept of campus network, not only allows agile deployment of data-driven applications and flexible offloading of computationally complex tasks, but also significantly accelerates the response times therein. This reduction in latency is particularly important for real-time DT applications such as collaborative telepresence and cobots.

### D. SENSING AND POSITIONING
Equipped with powerful AI solutions and fueled with the strong computing capability of MEC, DT systems still require massive amounts of information, as accurate and timely as possible, to support the modeling of certain assets, objects, or humans. The position is commonly, if not always, a key and essential part of this information, especially in mobile and flexible environments where systems and subsystems may change their position over time. In these cases, accurate position information can greatly improve the performance of DTs by helping them understand their context and environment, so that the DTs can automatically estimate the new location of these systems to adapt the applications accordingly.

The emerging technologies of Integrated Sensing and Communication (ISAC) exhibit great potential in improving the localization performance, in both accuracy and power efficiency, regardless if the target systems are capable of transmitting and/or receiving messages [50]. As another enabler, sensor fusion is known to be effective in enhancing the localization accuracy of the same target by merging the location information from different sources, e.g., barometric and radio sensors [51]. This principle is also well supported by the DT framework that helps to aggregate, broker, and exploit such information. Furthermore, with its freedom in manipulating wireless signals and marking the wireless channel, reconfigurable intelligent surface (RIS) has raised research interests regarding its application in indoor positioning to bring a significant accuracy gain at a low cost [52].

### E. HUMAN-MACHINE INTERFACE
In addition to the physical status of physical objects and human bodies, such as position, speed, temperature, etc., which can be straightforwardly measured by modern sensors, the future industrial DT system shall also be aware human mental status, in order to work human centric. The mental status information of human participants is particularly important in industrial processes, not only because of its correlation to the working efficiency, but more importantly, regarding the safety of the system and humans themselves. Recent developments in novel human-machine interface allows us to detect critical mental status (especially comprehension, concentration, fatigue, and emotion) from the directly observable physical features (including speech voice, facial expressions, galvanic skin response signal, eye movements, and bioelectric signals). With such mental status information integrated, human DT can be evolved to better understand humans and predict their behaviors, which will play a key role in human-presence-aware industries and cobot applications.

Besides capturing human status information, novel HMI with multi-sensory feedback technologies will also create new possibilities for human perception of information. Since vision and hearing are the most important human senses, visual and auditory user interfaces (UIs) will still indefinitely remain the dominating approach for industrial systems sending information to humans. However, the traditional visual/auditory UIs based on text, two-dimensional graphics and audio cannot fulfil the requirements of future industrial DT applications like MR, immersive telepresence, and human-robot collaboration, and therefore must be extended. The deployment of holographic vision, tactile feedback, and even neural stimulation technologies [53], will not only open a new door to highly efficient interaction between humans and DT-driven machines, but also make it possible for challenged people to participate in industrial processes.

### F. COMMUNICATION-COMPUTATION-CONTROL CODESIGN (COCOCOCO)
The massive deployment of DT in future industry is leading to complex coupling and interactions among different networked control systems (NCS), each with its individual modules of communication, computation, and control. Limited resources (e.g., radio resources and CPU circles) must be shared to a certain extent, not only among different NCS, but also among different modules of the same NCS. Such complex inter-module and inter-system dependency is challenging the classical paradigm of system design, and calling for a new framework of communication-computation-control codesign (CoCoCoCo), which shall help not only maximize the resource efficiency, but also deliver a guaranteed dependability in massive twinning for future industrial systems.

## VI. CONCLUSION
The adoption of DT technology in communication networks has garnered substantial attention for several years. The telecommunications community is largely in agreement that DT technology has the ability to revolutionize the 6G communication systems. Our exploration in this survey has revealed to us the great potential of DT technology on supporting the connectivity of I4.0 over the future 6G mobile communication networks. On the one hand, with the convenience in monitoring, simulation, and controlling provided by digital twins, as well as the ubiquitous, massive, and reliable radio access supported by 6G, it sketches a blueprint for an emerging ecosystem of industrial DTs, which interconnects everything and everybody, therewith greatly releases the potentials of DT technology and maximizes its implication in future industrial systems. On the other hand, the deployment of DT technologies is also enabling various emerging application scenarios of 6G, and enhancing the performance of 6G infrastructure itself. Nevertheless, towards this vision, plenty technical gaps are still to be bridged, and further investigation on the key enabling technologies are expected.

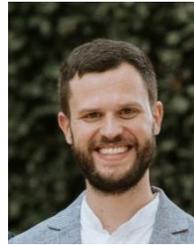

**BJOERN RICHERZHAGEN** obtained his M.Sc. degree in electrical engineering and information technology from Technische Universität Darmstadt, Germany, in 2012. From 2013 till 2019, he worked as research assistant and, from 2017 on, as research group head for the distributed sensing systems group at the Multimedia Communications Lab at TU Darmstadt, where he obtained his Ph.D. degree. Since 2019 he is with Siemens Technology, the central research and development division of Siemens. His research interests include adaptivity and self-organization in (heterogeneous) communication systems, P2P and mesh networks, and industrial features in wireless networks, especially 5G and 6G.

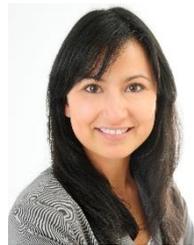

**KIM SCHINDHELM** graduated from the Ludwig-Maximilians Universität München in 2008 with a Diplom in computer science. After research activities at Siemens corporate research in Princeton, USA, and at LMU, Munich, she joined as a research scientist at Siemens Technology, the central research branch within Siemens AG. Her research activities cover wireless network architectures (especially regarding localisation and sensing), localisation algorithms, and semantic internet of things in the domains of industry, smart infrastructures and healthcare.

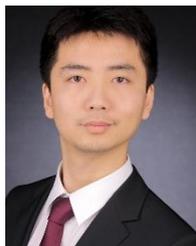

**BIN HAN** (Senior Member, IEEE) received his B.E. degree in 2009 from Shanghai Jiao Tong University, M.Sc. in 2012 from Technical University of Darmstadt, and the Ph.D. degree in 2016 from Karlsruhe Institute of Technology. Since July 2016 he has been with the Division of Wireless Communications and Radio Positioning, RPTU Kaiserslautern-Landau (formerly: Technical University of Kaiserslautern) as Postdoctoral Researcher and Senior Lecturer. His research interests are in the broad area of wireless communication and networking, with current focus on B5G/6G and MEC. He is the author of one book, five book chapters, and over 50 research papers. He has participated in multiple EU FP7, Horizon 2020, and Horizon Europe research projects. Dr. Han is Editorial Board Member for Network and Guest Editor for Electronics. He has served in organizing committee and/or TPC for IEEE GLOBECOM, IEEE ICC, EuCNC, European Wireless, and ITC. He is Voting Member of the IEEE Standards Association Working Groups P2303 and P3106.

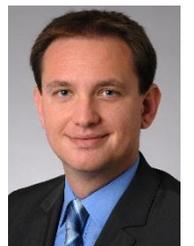

**FLORIAN ZEIGER** holds a PhD in the area of Networked Robotics. Since more than 15 years he is active in research and technology transfer in the areas of wireless sensor networks, ad-hoc networks, remote operation and control of mobile platforms, as well as Internet of Things. He was working as Research Assistant at the Department of Robotics and Telematics at the University of Würzburg until 2008. From 2008 to 2011 he worked as head of the department Communication and Remote Control at the research institute Zentrum für Telematik e.V. From 2011 to 2014 he worked as Senior Researcher and Project Manager at AGT International. In August 2014 he joined Siemens AG and he is now working as Portfolio Project Manager for Deterministic Communication & Edge Networks. Florian Zeiger was Project Manager, Technical Coordinator, and Key Expert in several large-scale national and international public funded R&D projects, as well as in industrial research and technology transfer projects. He is a certified PMI Project Management Professional, as well as a Certified SCADA Security Architect (IACRB).

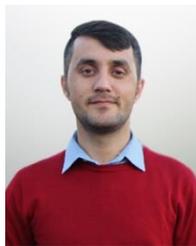

**MOHAMMAD ASIF HABIBI** received his B.Sc. degree in telecommunication engineering from Kabul University, Afghanistan, in 2011. He obtained his M.Sc. degree in systems engineering and informatics from the Czech University of Life Sciences, Czech Republic, in 2016. Since January 2017, he has been working as a research fellow and Ph.D. candidate at the Division of Wireless Communications and Radio Positioning, RPTU Kaiserslautern-Landau (formerly: Technical University of Kaiserslautern), Germany. From 2011 to 2014, he worked as a radio access network engineer for HUAWEI. His main research interests include network slicing, network function virtualization, resource allocation, machine learning, and radio access network architecture.

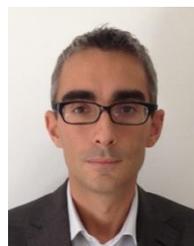

**FABRIZIO LAMBERTI** (Senior Member, IEEE) received his M.Ss. and his Ph.D. degrees in computer engineering from Politecnico di Torino, Italy, in 2000 and 2005. He is now a full professor at the Department of Control and Computer Engineering of Politecnico di Torino. His research interests include computer graphics, human-machine interaction and intelligent systems. He is serving as Associate Editor for IEEE Transactions on Computers, IEEE Transactions on Learning Technologies, IEEE Consumer Electronics Magazine, and the International Journal of Human-Computer Studies. He is also serving as Senior Associate Editor for IEEE Transactions on Consumer Electronics. He is a Member of the BoG of IEEE Consumer Technology Society (2021-23 term), for which he is also serving as VP of Technical Activities.

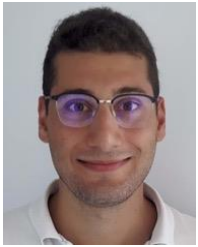
**FILLIPO GABRIELE PRATTICÒ** (Member, IEEE) received the M.Sc. degree in computer engineering from Politecnico di Torino, Turin, Italy, in 2017. Currently, he is a post-doc research assistant at the Department of Control and Computer Engineering of Politecnico di Torino, where he carries out research in the areas of extended reality, human-machine interaction, educational and training systems, and user experience design.

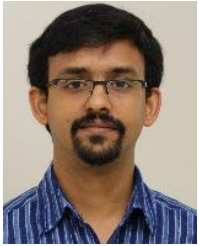
**KARTHIK UPADHYA** (Member, IEEE) is a senior radio research specialist at Nokia Bell Labs, Espoo, Finland. He received the M.Tech. degree from IIT Madras, India, in 2011 and DSc. (Tech) from Aalto University, Finland in 2018. He was a Visiting Researcher with the Wireless Networking and Communications Group, The University of Texas at Austin in 2017. His research interests include signal processing, massive MIMO and physical layer security.

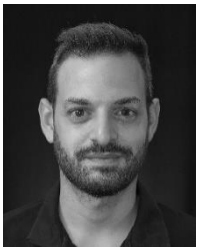
**CHARALAMPOS KOROVESIS** has been working as a professional 3D Artist and modeler since 2016. He is a Production and Management Engineer and graduated from Technical University of Crete in 2011. He attained a Master in Digital Arts from the Athens School of Fine Arts at 2015. He is currently working on the development of Augmented Reality (AR), Virtual Reality (VR) and Digital Twin applications for various EU projects among which 5G Tours, DEDICAT 6G and Hexa-X project. His main interests lie in the creative fields of photography, design and 3D graphics.

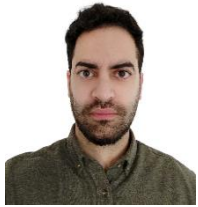
**IOANNIS-PRODROMOS BELIKAIDIS** is an experienced engineer at WINGS ICT Solutions, Athens Greece. He received his Bachelor and M.Sc. in Electrical and Computer engineering and his Ph.D. in Telecommunications and Digital Systems. His research interests include algorithms and procedures for complex networks, mobile pervasive and distributed computing, orchestration, management, and network slicing of resources. He has been involved in many EU funded projects, among which SPEED-5G, ONE-5G, Fantastic-5G, 5G-EVE, 5G-TOURS and is currently working in Hexa-X project.

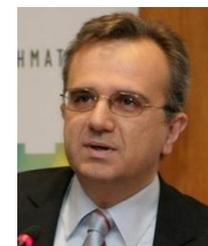
**PANAGIOTIS DEMESTICHAS** (Senior Member, IEEE) is a Professor at the University of Piraeus, School of ICT, Department of Digital Systems, Greece. Currently, he focuses on the development of systems for WINGS ICT Solutions (www.wings-ict-solutions.eu) and its spin-out Incelligent (www.incelligent.net). WINGS focuses on advanced solutions, leveraging on IoT / 5G / AI / AR, for the environment (air quality), utilities and infrastructures (water, energy, gas, transportation, construction), production and manufacturing (aquaculture, agriculture and food safety, logistics and industry 4.0), service sectors (health, security). Incelligent focuses on products for banking, the public sector and for telecommunication infrastructures. Panagiotis conducts research on 6G, cloud and IoT, big data and artificial intelligence, orchestration / diagnostics and intent-oriented mechanisms. He holds a Diploma and a Ph.D. degree on Electrical Engineering from the National Technical University of Athens (NTUA). He holds patents, has published numerous articles and research papers, and is a member of the Association for Computing Machinery (ACM) and a Senior Member of IEEE.

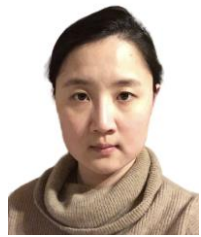
**SIYU YUAN** received the B.S. degree in electrical engineering from Qingdao University, Qingdao, China in 2015 and the M.S. degree in Automation and Robotics from TU Dortmund, Dortmund, Germany, in 2019. Since 2020 she has been working in the Division of Wireless Communications and Radio Positioning at RPTU, and is currently pursuing the Ph.D. degree in communication engineering. Her research topics include swarm intelligence with communication, simultaneous localization and mapping in approach of simulation, and processing of 3D or 2D images combined with machine learning method.

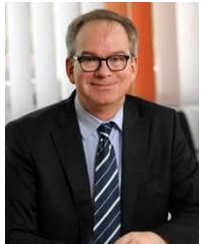
**HANS D. SCHOTTEN** (Member, IEEE) received the Ph.D. degree from the RWTH Aachen University, Germany, in 1997. From 1999 to 2003, he worked with Ericsson. From 2003 to 2007, he worked with Qualcomm. He became the Manager of a R&D Group, a Research Coordinator for Qualcomm Europe, and the Director for Technical Standards. In 2007, he accepted the offer to become the Full Professor with the Technical University of Kaiserslautern. In 2012, he became a Scientific Director of the German Research Center for Artificial Intelligence (DFKI) and the Head of the Department for Intelligent Networks. He served as the Dean of the Department of Electrical Engineering, Technical University of Kaiserslautern from 2013 until 2017. He has authored more than 200 papers and participated over 40 European and national collaborative research projects. Since 2018, he has been the Chairman of the German Society for Information Technology and a Member of the Supervisory Board of the VDE.